\documentclass[twocolumn,showpacs,aps,prb,floatfix,superscriptaddress]{revtex4}
\usepackage{graphicx}
\usepackage{epsfig}
\usepackage{amsmath} 
\newcommand{\sumnear}{\mathop{\sum}_{\langle i j \rangle}}
\begin{document}
\title{Supersolid phase transitions for hardcore bosons on a triangular lattice}
\author{Xue-Feng Zhang }
\email{zxf@physik.uni-kl.de} \affiliation{Physics Dept.~and Res.~Center
OPTIMAS, Univ.~of Kaiserslautern, 67663 Kaiserslautern, Germany}
\affiliation{Institute of Theoretical Physics, Chinese Academy of
Sciences, P.O. Box 2735, Beijing 100190, China}
\author{Raoul Dillenschneider}
\affiliation{Physics Dept.~and Res.~Center
OPTIMAS, Univ.~of Kaiserslautern, 67663 Kaiserslautern, Germany}
\author{Yue Yu }
\affiliation{Institute of Theoretical Physics, Chinese Academy of
Sciences, P.O. Box 2735, Beijing 100190, China}
\author{Sebastian Eggert}
\affiliation{Physics Dept.~and Res.~Center
OPTIMAS, Univ.~of Kaiserslautern, 67663 Kaiserslautern, Germany}

\begin{abstract}
Hard-core bosons on a triangular lattice
with nearest neighbor repulsion are a prototypical example of 
a system with supersolid behavior on a lattice.
We show that in this model the physical origin of the supersolid phase
can be understood quantitatively and analytically by 
constructing quasiparticle excitations of defects that are
moving on an ordered background.  The location of the solid to supersolid 
phase transition line is predicted from the effective model 
for both positive and negative (frustrated) hopping parameters.
For positive hopping parameters the calculations agree 
very accurately with numerical quantum 
Monte Carlo simulations.  The numerical 
results indicate that the supersolid to
superfluid transition is first order.

\end{abstract}



\maketitle

\section{Introduction}

A supersolid is characterized by a non-trivial solid order parameter 
which co-exists with an off-diagonal long-range superfluid order
parameter.\cite{supersolid1}  This state appears counter-intuitive since
the superfluid density implies ballistic coherent transport without disturbing the 
ordered background of the solid.
The possibility of a supersolid was discussed in the early 1970's and since then
much attention has been focused on $^4$He as the most likely experimental 
realization, which remains controversial.\cite{supersolid_experiment1}  
The motions of 
many different kinds of defects have been 
considered as possible mechanisms for supersolidity in $^4$He,
including
vacancies,\cite{supersolid1} grain boundaries
\cite{grain_boundary} or dislocations
 \cite{dislocation} which may be able to move through a solid state.
In recent years supersolid behavior on lattices has been 
established theoretically in a number of interacting boson 
models,\cite{wess,tri2,tri4,tri6,tri7,tri8,tri9,imp} which now gives hope
for an experimental realization of a supersolid with the help 
of ultra-cold gases.\cite{sengstock}
One prototypical example to exhibit a supersolid
phase are 
hard-core bosons with nearest neighbor repulsion on a triangular lattice 
\cite{GL,wess,tri2,tri4,tri6,tri7,tri8,tri9} which are described by the Hamiltonian
\begin{eqnarray}
H_0&=&
-t\sumnear(b_{i}^{\dag}b_{j}+b_{j}^{\dag}b_{i})+
V\sumnear n_{i}n_{j} -\mu\mathop{\sum}_in_i, \label{BH}
\end{eqnarray}
where $b_{i}^{\dag}$ $(b_i)$ is the boson creation (annihilation)
operator, $t$ describes the hopping amplitude, $\mu$ is the
chemical potential, $V$ reflects the repulsion effect, and $\langle
i,j\rangle$ represents nearest-neighbor sites.

For positive hopping $t>0$ a supersolid phase has been shown
using numerical simulations \cite{wess,tri2,tri4} and variational
wavefunctions.\cite{tri6,tri7}  For negative hopping $t<0$
it is not possible to use quantum Monte Carlo (QMC) simulations, but at least for the
special case of half filling a supersolid phase has been 
confirmed.\cite{tri8,tri9}  While the existence of the supersolid phase has been firmly 
established in this model,  it would be desirable to obtain a better understanding of the 
physical mechanism leading to the supersolid phase and an 
analytical description of the quantum phase transition.

\begin{figure}
\includegraphics[width=0.95\columnwidth]{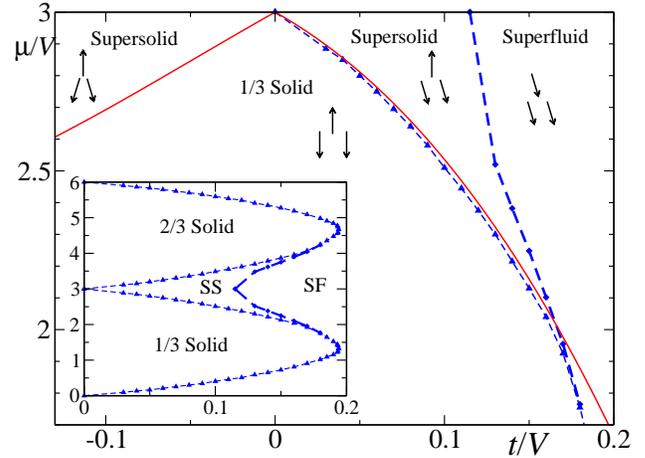}%
\caption{ The phase diagram of hard-core bosons
on the triangular lattice. The dashed lines (blue) are
numerical QMC results while the   
solid lines (red) are the analytical 
results from Eqs.~(\ref{cl}) and (\ref{cl2}).  The arrows indicate the order in terms of
the classical directions in the spin-wave theory. 
Inset: Overview of the entire particle-hole symmetric region.
 \label{pd}}
\end{figure}

The phase diagram of the model is shown in Fig.~\ref{pd}.  
For $0<\mu<3V$ and $t=0$ the solid
order is described by a 1/3-ordered state, 
where the unoccupied sites form a honeycomb sublattice around the filled sites
as indicated in Fig.~\ref{hop}.   
Due to particle-hole symmetry
the equivalent behavior can be seen for $\mu/V > 3$ in the form of a 2/3 filled state
(see inset of Fig.~\ref{pd}).
The goal of this paper is an analytic quantitative description of the transition 
to the supersolid phase which occurs for $t\neq 0$ around $\mu/V=3$.
The solid order is three-fold degenerate according to the choice of sublattice, 
however this degeneracy is {\it not} important for the formation of the supersolid state
so we 
will only consider one representative $1/3$-filled state in what follows.
Finite hopping $t\neq 0$ causes quantum fluctuations in form of
virtual excitations from the occupied sites onto the
unoccupied ones and back. 
Therefore the ordered state is renormalized, e.g. the density
of the unoccupied honeycomb lattice is now finite $\sim 3 t^2/4V^2$ to lowest order
and the occupied density is reduced $\sim 1-3 t^2/2V^2$.
However,  the delocalization of these virtual excitations is not 
the driving mechanism for the transition into 
the supersolid state.  
Instead we conjecture that a different type of excitation
starts to appear at the phase transition, namely ''defects''
consisting of additional occupied sites.
These defects can only move on the 
honeycomb lattice of unoccupied sites, while the solid order remains 
surprisingly stable, which leads to an effective model.

\section{Effective model}
\begin{figure}
\includegraphics[width=0.95\columnwidth,clip]{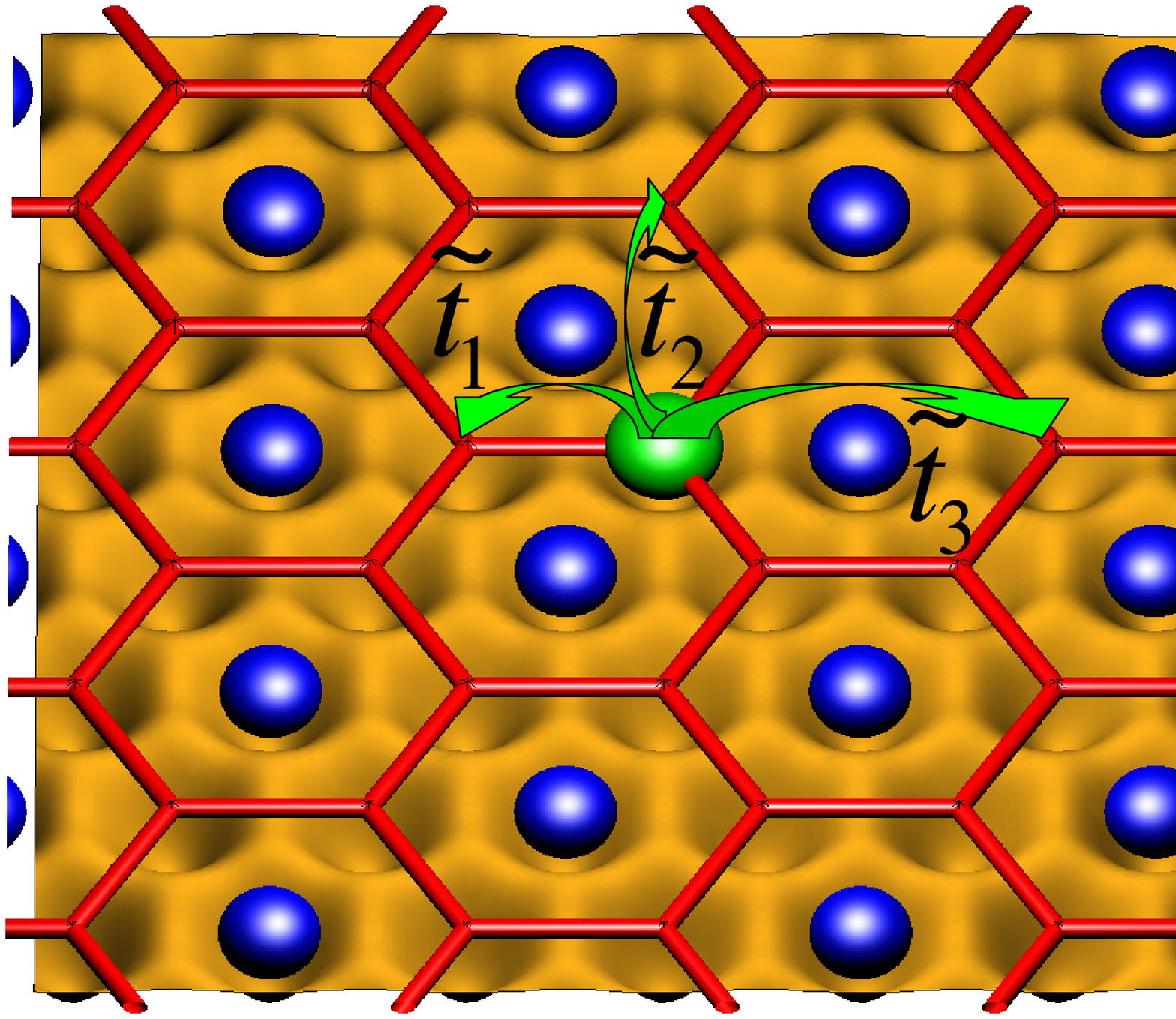}%
\caption{ 
Schematic description of the effective model 
on the unoccupied honeycomb sublattice (lines)
where the  defect (large central sphere) can hop to neighboring sites.
Smaller spheres (blue) are occupied sites, which 
only participate in virtual excitations.
 \label{hop}}
\end{figure}

The corresponding effective Hamiltonian for hard-core boson defects
on the honeycomb lattice
can be derived in the strong coupling expansion.
Ordinary perturbation theory has a diverging number 
of contributions, 
but it is possible to analyze the movement of a single defect by only taking those terms
into account which actually involve the defect.\cite{schmidt}
This results in the following effective Hamiltonian on the honeycomb sublattice
\begin{equation}
H_{\rm eff} = -\sum_{i,d} \tilde{t}_{d} (a_i^\dagger a_{i+d}^{\phantom{\dagger}}
+a_{i+d}^\dagger a_i^{\phantom{\dagger}})
+V \sumnear m_{i} m_j - \tilde{\mu} \sum_i m_j, \label{Heff}
\end{equation}
where $m_i = a_i^\dagger a_i^{\phantom{\dagger}}$ and
the effective 
hopping parameters are 
given by $\tilde{t}_1 = t +t^2/V$
for nearest neighbors,
$\tilde{t}_2 = t^2/V$
for next nearest neighbors,
and $\tilde{t}_3 = t^2/V$ 
for next-next nearest neighbors 
on the honeycomb sublattice as shown in Fig.~\ref{hop}. 
The second-order terms
arise from a two-step hopping process, where a boson from the occupied sublattice
jumps on an unoccupied site and then the defect takes the place of the original boson, 
thereby effectively transporting the defect.
Likewise, the effective chemical potential is changed
 $\tilde{\mu} = \mu -3V + 5 t^2/2V$ since
virtual excitations from the unoccupied sublattice and back are
now affected by the defect, which results in an effective self-energy. 
Since the  defects can move ballistically and coherently on the unoccupied
honeycomb sublattice according to the effective model, 
a finite density of such defects also explains the 
coexistence of a finite superfluid density on top of a relatively stable renormalized 
ordered state, which are in fact the central characteristics of the supersolid phase.

The effective model in Eq.~(\ref{Heff}) is of course still an interacting hard-core boson 
model, but is 
much easier to treat since we only have to consider small densities of defects as
will be seen below.
In particular, according to our hypothesis 
the phase transition line is simply determined by the point when it is
energetically favorable to create one single defect.
For positive hopping $t>0$ the energy of a single defect is minimized
in the uniform $k=0$ state with a total energy of 
$\tilde{E}_{+} = -3 \tilde{t}_1 - 6\tilde{t}_2 -3 \tilde{t}_3 -\tilde{\mu}$, which
must change sign at the phase transition line.
By taking also third order processes into account this results in the following
prediction for the 
phase transition line between solid and supersolid
\begin{eqnarray}
\mu_{+}&=&3V-3t-\frac{29t^2}{2V}-\frac{77t^3}{4V^2} +{\cal{O}}(t^4). \label{cl}
\end{eqnarray}
As shown in Fig.~\ref{pd} this result agrees remarkably well with our 
QMC simulations, where we used the stochastic series expansion
with parallel tempering.\cite{sse1}  The errorbars for the data from the QMC simulations
are smaller than the size of the symbols in all figures.
For $\mu > 3V$ the particle hole symmetric line 
is given by $\tilde{\mu}_+ = 6V - \mu_+ $.

A similar analysis can be made for negative hopping $t<0$. 
Since the honeycomb lattice is bi-partite the 
positive nearest neighbor hopping can be restored by 
a simple Gauge transformation of $-1$ on one sublattice, but the next-nearest
neighbor coupling remains negative. Therefore, 
the ground state energy of one defect is now 
$\tilde{E}_{-} = 3 \tilde{t}_1 - 6\tilde{t}_2 +3 \tilde{t}_3 -\tilde{\mu}$,
and accordingly we obtain a phase transition line of the form
\begin{eqnarray}
\mu_{-}&=&3V+3t-\frac{5t^2}{2V}-\frac{71t^3}{4V^2} +{\cal{O}}(t^4), \label{cl2}
\end{eqnarray}
where we have again taken third order processes into account.
For negative hopping it is more difficult to treat the problem
with numerical methods, so there are no independent checks for this prediction so far.
However, considering the good agreement for the positive hopping case above, it is reasonable 
to expect that the same calculation also applies for negative hopping with similar 
accuracy.  We therefore find that the supersolid region is stable but 
slightly narrower for negative hopping, which 
appears to be caused by the additional kinetic frustration.

\section{analytical and numerical results near the phase transition}
In order to calculate the detailed behavior near
the phase transition analytically it is now possible to 
use linear spin-wave theory on the effective Hamiltonian in Eq.~(\ref{Heff}).  
It should be noted that the 
system is at low filling (i.e.~nearly saturated in terms of spins)
 and that the honeycomb lattice has two nonequivalent sites, but otherwise
the spin-wave calculation follows the standard scheme.\cite{Weihong}
In particular, 
the hard-core bosons can be mapped exactly onto spins, which are expressed in a 
rotated frame according to the classical alignment.  The classical angle relative to the 
$z$-direction is given by $\cos\theta =B/(3 V/2-\sum_d \tilde{t}_d$) in terms of the field $B=\tilde \mu -3V$.
The spins are then re-expressed in terms of regular bosons, which has the advantage
that the interaction can be solved by a Bogoluibov transformation to linear order. 
This procedure is especially reliable in the low density limit near the
phase transition that we are interested in.
The details of this calculation can be found in the appendix.
The density $\langle\rho\rangle = \frac{1}{2}-\langle S^z\rangle$ on the honeycomb sublattice 
at $T=0$ is then given in terms of the Bogoluibov rotation angles
\begin{equation}
\langle\rho\rangle =  
\frac{1}{2} - \cos \theta \left[\frac{1}{2}
- \frac{2}{N} \sum_k \left(\sinh^2 \theta_k^\alpha + \sinh^2 \theta_k^\beta\right) \right],
\label{sw-dens}
\end{equation}
\begin{tabular}{ll}
where & $\tanh 2 \theta_k^\alpha = - |\gamma_k| A /\left(t +|\gamma_k|\left(A - t \right)\right)$  \\
and &  $\tanh 2 \theta_k^\beta = - |\gamma_k| A /\left(t -|\gamma_k|\left(A - t \right)\right)$\\
with & $A = \sin^2 \theta \left(2t + V \right)/4 $ and \\
 $|\gamma_k| =$ & $
\frac{1}{3} \left[3 + 4 \cos \frac{3 k_x}{ 2}  \cos \frac{\sqrt{3} k_y}{2} 
+ 2 \cos \left(\sqrt{3} k_y \right)\right]^{1/2} $. \\
\end{tabular}

\begin{figure}
\includegraphics[width=0.95\columnwidth]{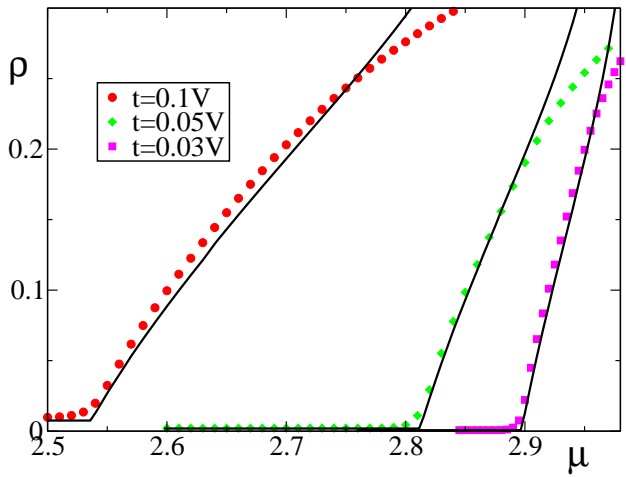}%
\caption{ The density on the sites with lower filling on the 
triangular lattice near the phase transition (symbols) for $t=0.03V$ ($V\beta = 300$), $t=0.05V$ ($V\beta=150$)
and $t=0.1V$ ($V\beta=100$).  The solid lines are the analytical prediction from Eq.~(\ref{sw-dens}).
 \label{dens}}
\end{figure}
\begin{figure}
\includegraphics[width=0.95\columnwidth]{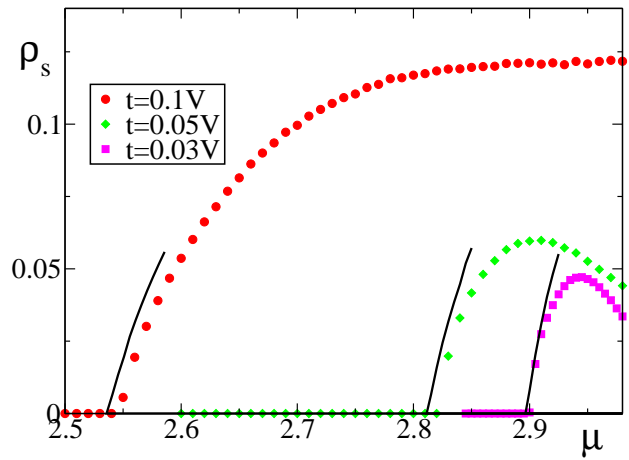}%
\caption{ The superfluid density on the
triangular lattice near the phase transition (symbols).
The solid lines are the analytical prediction from Eq.~(\ref{sw-rhos}).
 \label{rhos}}
\end{figure}

In Fig.~\ref{dens} we show the density on the unoccupied sublattice from QMC simulations on 
a triangular lattice with $24^2$ sites near the solid to supersolid phase transition at 
different hoppings (symbols).
This can be compared to the spin-wave results for the density on the honeycomb lattice 
from Eq.~(\ref{sw-dens}) (lines), where   
the small background contribution $3 t^2/4V^2$ from virtual 
excitations was added.  The virtual excitations are localized so this finite 
background density does not contribute to the superfluid density which is also 
confirmed by the data in Fig.~\ref{rhos}.
Here and in what follows the spin-wave results always 
include the additional third order shift in $\mu$ by $77 t^3/4V^2$ from Eq.~(\ref{cl}).  
The good agreement  not only 
confirms the validity of the effective model on the honeycomb 
lattice in Eq.~(\ref{Heff}), but also provides an analytical 
tool to calculate the general behavior and correlations near the phase transition 
for both positive and negative hopping $t$.
As a check we have also computed densities and correlations on the honeycomb lattice,
which agree equally well with the original triangular model (not shown).
The correlations
between the occupied sublattice and the unoccupied honeycomb sublattice
can be calculated by the modified perturbation theory discussed above, which 
results in positive $\langle b^\dagger_i b_j\rangle$ correlations between sublattices 
for $t>0$,  while the correlations cancel
to second order for $t<0$, i.e. the occupied sublattice always gives only a very weak
contribution to the superfluid density as indicated by the spin directions in Fig.~\ref{pd}.

The superfluid density on the unoccupied honeycomb sublattice can be 
calculated in the effective spin model as 
the response to a small twist $\nu$ 
in the x-y plane $\rho_s = d^2 \langle H(\nu) \rangle / d^2 \nu$. 
Expanding
the Hamiltonian leads to $H(\nu) = H + \sum_{ i,d } \left(\nu_{d} J^{s}_{d}
+ \nu_{d}^2/2 T_{d} \right)$, where $\nu_{d}$ is the projection of the twist along the 
corresponding bond direction, 
$J^{s}_{d} = (i/2) \widetilde{t}_{d} \left( S_i^{+}S_{i+d}^{-}
- S_{i+d}^{+}S_i^{-} \right)$ is the spin current, and 
$T_{d} = (1/2) \widetilde{t}_{d} \left( S_i^{+}
S_{i+d}^{-} + S_{i+d}^{+}S_i^{-}\right)$
is the spin-kinetic energy.\cite{Schulz,Cuccoli}  
In  linear spin-wave theory the response to 
the spin current does not contribute in the low density limit and the 
superfluid is therefore approximated by the correlation functions 
$T_d$ of neighbors connected by $\tilde{t}_d$ 
\begin{equation}
\rho_s \approx  \frac{1}{2} (  
T_1+4T_2+3T_3)
\label{sw-rhos}
\end{equation}
which can be calculated in the rotated spin frame (see appendix).
In Fig.~\ref{rhos} we compare the results from Eq.~(\ref{sw-rhos}) on the honeycomb 
lattice with the superfluid density $\rho_s$ derived from the winding
number $\langle W^2\rangle/4 \beta t$  in QMC simulations for the triangular lattice 
in the thermodynamic limit near the phase transition.
The predictions from the effective model also show a sharp increase
at the phase transition, but the agreement is limited to a smaller region since the
contribution of the spin current response was not included.
In any case, the agreement near the phase transition line again gives  strong support for
the validity of the effective model in Eq.~(\ref{Heff}).

\section{Analysis of the phase transition}

\begin{figure}
\includegraphics[width=0.95\columnwidth]{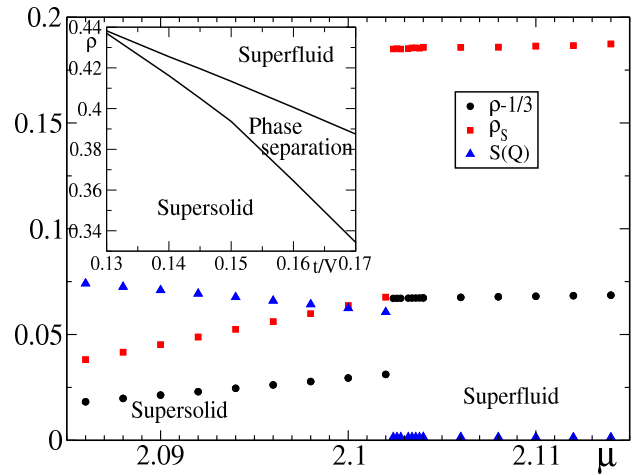} 
\caption{ 
The density, structure factor $S(Q) 
=\langle
|\mathop{\sum}_{r} \hat S^z_r e^{i Q \cdot r} |^2\rangle/N^2$ at $Q=\!\![4/3\pi,0]$
and the
superfluid density with $L=36$ and $T=0.01$ at $t=0.16$ and $V=1$
by varying $\mu$.  
Inset: Observed region of phase separation between supersolid and superfluid phases.
 \label{jump}}
\end{figure}

Finally, we would like to consider the implications of our model for the 
supersolid to superfluid transition, which has been reported to be second order
before.\cite{wess}  If our model is correct this transition should simply 
correspond to a melting of the solid due to the appearance of more
and more virtual excitations, i.e.~a breakdown of the perturbation expansion.
While we have not found an energetic condition which would quantify the location of the
supersolid to superfluid transition line, we believe that such a melting transition should be 
first order.
One reason is also that a third order term in the
Ginzburg Landau expansion is generically 
allowed by momentum conservation for an order parameter with
triangular symmetry, so that a continuous transition can only occur at special 
points.\cite{rosch}
We have therefore tested the behavior of the phase transition with QMC
simulations.  In Fig.~\ref{jump} it can be seen that for $t=0.16$
the structure factor
abruptly goes to zero close to $\mu =2.012(2)$ while the densities show
clear discontinuities.  For larger $\mu$ and smaller $t$ 
closer to the particle-hole symmetric point these discontinuities become systematically 
smaller, but can still be observed even without histogram methods.
The discontinuity in $\rho$ corresponds to a region of phase separation if the 
density is held fixed as shown in the inset.
During preparation of the paper we discussed our
discovery of the first order behavior with one of the authors of Ref.~[\onlinecite{wess}]
and since then two quantitative studies have appeared which 
confirm the first order behavior.\cite{wessnew}

\section{Conclusions}

In conclusion we proposed an analytical model in the form of
defects moving in an ordered background, which gives a quantitative description of the
solid to supersolid transition in the hard-core boson model.  In this
way the transition line can be determined and correlations nearby 
can be calculated  with the help of spin-wave theory for positive and negative hopping $t$.
For positive $t$ the calculations agree well with QMC simulations.
 For $t<0$ we make a quantitative prediction for the supersolid region, 
which has so far not been estimated by numerical methods.
We find that the kinetic frustration for $t<0$ does not destroy 
the supersolid phase, but the higher order corrections imply a narrower region.  
 In fact, the higher order terms imply a possible maximum of the extent of the
supersolid phase  with increasing $|t|$ in 
this case.  It is indeed feasible that a 
configuration corresponding to a low density (spin down) on one sublattice 
and a small x-y antiferromagnetic order on the other two sublattices may be stable
all the way to the SU(2) invariant point $V=-2t=\mu/3$.
In this case the SU(2) invariant point would play the role of a trivial 
''degenerate'' supersolid where a superposition of broken orders is 
possible in any direction.

\begin{acknowledgments}
The authors would like to thank Y.C. Wen, A. Rosch, S. Wessel,  A. Paramekanti, 
A. Struck and S. S\"{o}ffing for helpful discussions and suggestions.
This work was supported by the DAAD, and the DFG via the Research
Center Transregio 49.
\end{acknowledgments}

\appendix*

\section{Spin wave analysis}

\newcommand{\tr}{\text{Tr}}
\def\UnitMatrix{\mbox{1\hspace{-.25em}I}}

We consider a two dimensional system of hardcore bosons on an
effective hexagonal lattice in Eq.~(2). 
To enforce the hardcore constraint in a simple way, the Hamiltonian 
is mapped onto the two dimensional XXZ model with external magnetic field using
$a_i^{\dagger} \leftrightarrow S_i^{+}$, 
$a_i \leftrightarrow S_i^{-}$, and ${\hat n_i}\leftrightarrow
S_i^z+1/2$. With this mapping the hardcore boson model becomes
\begin{equation}
H = - \sum_{i,d} 
 \widetilde{t}_d (S_i^+ S_j^- + S_i^-S_j^+)   
+\sum_{\langle ij \rangle}  V S_i^zS_j^z - B \sum_i S_i^z
\label{EqXXZ1}
\end{equation}
where $B=\widetilde{\mu} - \lambda V$ is an effective field
and $\lambda = 3/2$ is half the coordination number
 on a hexagonal lattice.
Here and in what follows we generally omit the ground state energy.

The $z$-axis can be aligned  
with the mean field magnetization direction by the rotation
$S_i^x = S_i^{\prime x} {\rm cos}(\theta) + S_i^{\prime z} {\rm
sin}(\theta)$, $S_i^y = S_i^{\prime y}$ and 
$S_i^z = -S_i^{\prime x}{\rm sin}(\theta) + S_i^{\prime z}{\rm cos}(\theta)$,
where ${\vec S}_i^{\prime}$ is the spin vector in the rotated
frame.
The hexagonal lattice has two spins per unit cell, which are 
are expressed in terms of bosonic
operators $c_i$ ($c_i^\dagger$) and $d_i$ ($d_i^\dagger$) respectively, i.e.
$S_i^{\prime x}= \frac{1}{2}(S_i^{\prime +}+S_i^{\prime -}) = \frac{1}{2}
(c_i^{\dagger}+c_i)$, $S_i^{\prime y}= \frac{1}{2i}(S_i^{\prime +}-S_i^{\prime -}) =
\frac{1}{ 2{\rm i}}(c_i^{\dagger}-c_i)$, $S_i^{\prime z}= \frac{1}{2} - c_i^{\dagger}c_i$
and likewise for the other sublattice in terms of $d_i$ ($d_i^\dagger$).
Note that the spin-wave operators obey the usual (soft-core) bosonic 
commutation relations. 
Substituting these expressions into Eq.~\eqref{EqXXZ1} and assuming a
dilute gas of spin-waves, we can ignore cubic and quartic terms in the
bosonic operators. The linear term in the operators is required to vanish yielding,
$\cos \theta  = \left(  \widetilde{\mu} - 
\lambda V \right)/\left(\lambda V + \sum_{d} \widetilde{t}_d \right)$,
where the sum runs over the nearest, next-nearest and next-next-nearest neighbor bonds.
The longer range hopping terms of order $t^2$ play an 
important role in determining the mean field angle and
the superfluid density below, but they can be neglected 
when determining the Bogoluibov rotation.  Therefore,
the Hamiltonian reduces to
\begin{eqnarray}
H &=&
\sum_{<ij>}  \left\{ 
A \left(c_i^\dagger d_j^\dagger + c_i d_j \right)
+
\left( A - t\right) \left( c_i^\dagger d_j + c_i d_j^\dagger \right)  \right\}
\nonumber \\
&&
+3 t \left( \sum_{i \in A} c_i^\dagger c_i + \sum_{i \in A} d_i^\dagger d_i \right)
+H_{MF}, 
\label{EqHhexa}
\end{eqnarray}
\noindent
where $A = \sin^2 \theta \left(2t + V \right)/4$ and the mean-field Hamiltonian is given by 
\begin{equation}
H_{MF} = 
N\left[\frac{\cos^2\theta}{4}\left(\lambda V+\sum_d \tilde{t}_d \right)
 -\frac{\cos\theta}{2}B\right],
\end{equation}
which is minimized by the same classical condition $\cos \theta  = 
B/\left(\lambda V + \sum_{d} \widetilde{t}_d \right)$. 
Here $N$ is the total number of lattice sites.

The Hamiltonian \eqref{EqHhexa} can be diagonalized by  
the Fourier transformation $c_i = \sqrt{2/{N}} \sum_k e^{i \vec{k}.\vec{r}_i} c_k$
and $d_i = \sqrt{2/{N}} \sum_k e^{i \vec{k}.\vec{r}_i} c_k$.
We introduce the structure factor 
$\gamma_k = \frac{1}{3} \sum_d e^{i \vec{k}.\vec{r}_d} = |\gamma_k| e^{i \phi_k}$ 
which is a complex number. 
The phase $\phi_k$ can be absorbed in a gauge transformation 
$d_k \rightarrow e^{-i\phi_k} d_k$ in the Fourier transformed Hamiltonian
and the amplitude of the structure factor reads 
$|\gamma_k| = \frac{1}{3} \left[3 + 4 \cos \frac{3 k_x}{ 2}  \cos \frac{\sqrt{3} k_y}{2}
+ 2 \cos \left(\sqrt{3} k_y \right)\right]^{1/2} $.
The Hamiltonian can be diagonalized using the Bogoluibov transformation \cite{Weihong}
\begin{eqnarray*}
c_k &=& u^\alpha_k \alpha_k + v^\alpha_k \alpha^\dagger_{-k} + u^\beta_k \beta_k - v^\beta_k \beta^\dagger_{-k}
 \\
d_k &=& u^\alpha_k \alpha_k + v^\alpha_k \alpha^\dagger_{-k} - u^\beta_k \beta_k + v^\beta_k \beta^\dagger_{-k}
\end{eqnarray*}
\noindent
where 
$u^\alpha_k = \cosh \theta^\alpha_k$,
$v^\alpha_k = \sinh \theta^\alpha_k$,
$u^\beta_k = \cosh \theta^\beta_k$, and 
$v^\beta_k = \sinh \theta^\beta_k$, with
$\tanh{2 \theta^\alpha_k} = - |\gamma_k| A / \left(t +|\gamma_k|\left(A - t \right) \right)$ and 
$\tanh{2 \theta^\beta_k} = -|\gamma_k| A / \left(t -|\gamma_k|\left(A - t \right) \right)$.
Finally, the Hamiltonian \eqref{EqHhexa} takes the diagonal form
\begin{equation}
H= \sum_k \omega_k^\alpha \left( \alpha^\dagger_k \alpha_k + 1/2 \right)
+ \sum_k \omega_k^\beta \left( \beta^\dagger_k \beta_k + 1/2 \right).
\label{EqHDiag}
\end{equation}
\noindent
with $\omega_k^\alpha = 6 \sqrt{ ( t + |\gamma_k| (A-t))^2 - (A |\gamma_k|)^2}$ 
and $\omega_k^\beta = 6 \sqrt{ ( t - |\gamma_k| (A-t))^2 - (A |\gamma_k|)^2}$.


The boson density of the model in Eq.~(2) is directly given by
$\rho = \left(\langle S^z \rangle + 1/2 \right)$ where the magnetization reads
\begin{eqnarray*}
\langle S^z \rangle 
&=& \langle  - S_i^{\prime x}  \sin \theta + S_i^{\prime z} \cos \theta \rangle
\nonumber \\
&=& 
 \cos \theta \left[
\frac{1}{2}
- \frac{2}{{N}} \sum_k 
\left((v^\alpha_k)^2 + (v^\beta_k)^2 \right) 
\right].
\end{eqnarray*}
which follows from a straight-forward evaluation of the expectation values
in terms of the diagonal bosons in Eq.~(\ref{EqHDiag}).

The superfluid density
is given by the second derivative of the energy of the spin system
with respect to a uniform  
twist $\nu$ across the system $\rho_s = d^2 \langle H(\nu) \rangle / d^2 \nu$.
The Hamiltonian $H(\nu)$ is derived from the XXZ model in Eq.(\ref{EqXXZ1}) by application of a 
site dependent rotation by an angle $\nu_i$ around the $z$ axis,
$S_i^{+} \rightarrow S_i^{+} e^{i\nu_i}$, $S_i^{-} \rightarrow S_i^{-} e^{-i\nu_i}$
and $S_i^z \rightarrow S_i^z$. Expanding the Hamiltonian around $\nu_d = \nu_i -\nu_{i+d} = 0$
leads to $H(\nu) = H + \sum_{i,d } \left(\nu_{d} J^{s}_{d}
+ \nu_{d}^2/2 T_{d} \right)$, where $J^{s}_d = (i/2) \widetilde{t}_{d} \left( S_i^{+}S_{i+d}^{-} - S_{i+d}^{+}S_i^{-} \right)$
is the spin current and $T_{d} = (1/2) \widetilde{t}_{d} \left( S_i^{+}S_{i+d}^{-} + 
S_{i+d}^{+}S_i^{-}\right)$
is the spin-kinetic energy.\cite{Schulz,Cuccoli} 
To first-order in perturbation theory the spin stiffness
is given by 
$\rho_s = \frac{1}{2N} \frac{\partial^2}{\partial \nu^2} \sum_{i,d} \nu_{d}^2 \langle T_{d} \rangle$.
Second order perturbation leads to a term integrating the current-current correlator
with respect to $J_s$, which
can be neglected in our spin-wave approach.\cite{Cuccoli}
To obtain the spin stiffness a uniform twist $\nu$ along one direction is applied, 
which leads to
\begin{eqnarray}
\rho_s &=& \frac{1}{2N} \sum_{i,d} \left(\frac{\partial^2}{\partial \nu^2}  \nu_{d}^2 \right) 
\widetilde{t}_{d} \langle S_i^x S_{i+d}^x + S_{i+d}^y S_i^y \rangle,
\nonumber \\
&=&
\frac{\widetilde{t}_1}{2} \langle S_i^x S_{i+1}^x + S_{i+1}^y S_i^y \rangle
+
2\widetilde{t}_2 \langle S_i^x S_{i+2}^x + S_{i+2}^y S_i^y \rangle
\nonumber \\
&&
+
\frac{3\widetilde{t}_3}{2} \langle S_i^x S_{i+3}^x + S_{i+3}^y S_i^y \rangle
,
\end{eqnarray}
\noindent
The spin-spin correlation functions are provided by
$\langle S_i^x S_j^x + S_j^y S_i^y \rangle = \cos^2 \theta 
\langle S_i^{\prime x} S_j^{\prime x} \rangle +
\sin^2 \theta \langle S_i^{\prime z} S_j^{\prime z} \rangle
+ \langle S_i^{\prime y} S_j^{\prime y} \rangle$,
where $d=1,2,3$ denote nearest neighbor, next-nearest neighbor, 
and next-next-nearest neighbor, respectively. The spin-spin correlation functions
in the rotated frame read
%
%
{\small
\begin{eqnarray*}
\langle S_i^{\prime x} S_{i+d}^{\prime x} \rangle
&=&
\frac{2}{N} \sum_k f_d(\vec{k})
\Big(
(v^\alpha_k)^2 + u^\alpha_k v^\alpha_k
-
(v^\beta_k)^2 + u^\beta_k v^\beta_k 
\Big),
\nonumber \\
\langle S_i^{\prime y} S_{i+d}^{\prime y} \rangle
&=&
\frac{2}{N} \sum_k f_d(\vec{k})
\Big(
(v^\alpha_k)^2 - u^\alpha_k v^\alpha_k
- (v^\beta_k)^2 - u^\beta_k v^\beta_k
\Big),
\nonumber \\
\langle S_i^{\prime z} S_{i+d}^{\prime z} \rangle
&=& \left(\langle S^{\prime z} \rangle -1/4 \right), \forall d,
\end{eqnarray*}
}

\noindent
and the functions $f_d(\vec{k})$ are given by
%
%
%
%
\begin{eqnarray*}
f_{1}(\vec{k}) &=& 2|\gamma_k|,
\nonumber \\
f_{2}(\vec{k}) &=& 3|\gamma_k|^2-1
\nonumber \\
f_{3}(\vec{k}) &=& 
2|\gamma_k| + \frac{2}{9 |\gamma_k|}
\Big(
-3 + \cos(3 k_x)
\nonumber \\
&&
+ 2 \cos(3 k_x /2) \cos(3 \sqrt{3} k_y/2)
\Big)
.
\end{eqnarray*}

\bibliographystyle{apsrev}

\end{document}